\PassOptionsToPackage{unicode}{hyperref}
\PassOptionsToPackage{hyphens}{url}
\documentclass[
]{article}
\usepackage{xcolor}
\usepackage{amsmath,amssymb}
\usepackage{iftex}
\ifPDFTeX
  \usepackage[T1]{fontenc}
  \usepackage[utf8]{inputenc}
  \usepackage{textcomp} 
\else 
  \usepackage{unicode-math} 
  \defaultfontfeatures{Scale=MatchLowercase}
  \defaultfontfeatures[\rmfamily]{Ligatures=TeX,Scale=1}
\fi
\usepackage{lmodern}
\ifPDFTeX\else
\fi
\IfFileExists{upquote.sty}{\usepackage{upquote}}{}
\IfFileExists{microtype.sty}{
  \usepackage[]{microtype}
  \UseMicrotypeSet[protrusion]{basicmath} 
}{}
\makeatletter
\@ifundefined{KOMAClassName}{
  \IfFileExists{parskip.sty}{%
    \usepackage{parskip}
  }{
    \setlength{\parindent}{0pt}
    \setlength{\parskip}{6pt plus 2pt minus 1pt}}
}{
  \KOMAoptions{parskip=half}}
\makeatother
\usepackage{color}
\usepackage{fancyvrb}

\DefineVerbatimEnvironment{Highlighting}{Verbatim}{commandchars=\\\{\}}
\newenvironment{Shaded}{}{}

\newcommand{\AttributeTok}[1]{\textcolor[rgb]{0.49,0.56,0.16}{#1}}

\newcommand{\ExtensionTok}[1]{#1}

\newcommand{\NormalTok}[1]{#1}

\usepackage{longtable,booktabs,array}
\usepackage{caption}
\usepackage{calc} 
\usepackage{etoolbox}
\makeatletter
\patchcmd\longtable{\par}{\if@noskipsec\mbox{}\fi\par}{}{}
\makeatother
\IfFileExists{footnotehyper.sty}{\usepackage{footnotehyper}}{\usepackage{footnote}}
\makesavenoteenv{longtable}
\usepackage{graphicx}
\makeatletter
\newsavebox\pandoc@box
\newcommand*\pandocbounded[1]{
  \sbox\pandoc@box{#1}%
  \Gscale@div\@tempa{\textheight}{\dimexpr\ht\pandoc@box+\dp\pandoc@box\relax}%
  \Gscale@div\@tempb{\linewidth}{\wd\pandoc@box}%
  \ifdim\@tempb\p@<\@tempa\p@\let\@tempa\@tempb\fi
  \ifdim\@tempa\p@<\p@\scalebox{\@tempa}{\usebox\pandoc@box}%
  \else\usebox{\pandoc@box}%
  \fi%
}
\def\fps@figure{htbp}
\makeatother
\providecommand{\tightlist}{%
  \setlength{\itemsep}{0pt}\setlength{\parskip}{0pt}}
\usepackage[]{natbib}
\usepackage[margin=1in]{geometry}
\usepackage{xurl}
\usepackage{enumitem}
\usepackage{float}

\setlist{nosep}
\usepackage{bookmark}
\IfFileExists{xurl.sty}{\usepackage{xurl}}{} 
\makeatletter
\@ifundefined{xmpquote}{\newcommand{\xmpquote}[1]{#1}}{}
\makeatother
\hypersetup{
  pdftitle={Hints Help. But Do They Teach? Testing Skill Transfer in Code Generation},
  pdfauthor={Will Badr},
  pdfkeywords={\xmpquote{code generation}, \xmpquote{prompting}, \xmpquote{activation steering}, \xmpquote{resampling}, \xmpquote{correctness probing}},
  hidelinks,
  pdfcreator={LaTeX via pandoc}}

\title{Hints Help. But Do They Teach? Testing Skill Transfer in Code Generation}
\author{Will Badr\\\texttt{scpg0807@leeds.ac.uk}}
\date{}

\begin{document}
\maketitle
\begin{abstract}
When a hint turns a failing generated program into a passing one, does it supply missing information or help the model reach a solution it could already produce? We test these explanations on HumanEval+ and MBPP+ with executable evaluation. For Qwen2.5-3B-Instruct, an adaptive relevant-hint procedure rescues 36 of 79 selected failures; one unrelated hint rescues 19, while eight unhinted samples solve 46 and cover 31 of the 36 hint rescues. Phi-3.5-mini reproduces this pattern: 42 of 101 failures are rescued by relevant hints, 17 by an unrelated hint, and 57 by unhinted sampling, which covers 36 of the 42 rescues. Because the hint procedures use different attempt budgets, their difference does not isolate a semantic effect. Mechanistic tests on Qwen find a stable activation direction shared by relevant and unrelated hints. Persistent addition is associated with 14 rescues and 18 failures, with no detected net accuracy change, while the estimated advantage of learned low-rank interventions remains positive but imprecise. Full textual specifications solve 22 of 24 context-defined problems, compared with 5--11 for the tested virtual-KV prefixes. Post-generation hidden-state probes transfer across benchmarks with pooled AUROC 0.806 and 0.780, although their top-one selection advantage over token confidence is statistically unresolved. Overall, the adaptive relevant-hint procedure rescues failures under the implemented conditions, but most rescues are already accessible through sampling; the tested internal interventions do not establish task-general capability transfer.
\end{abstract}

\section{1. Introduction}\label{introduction}

A code model fails, receives a short hint, and passes. This before-and-after result is often described as capability transfer, but it does not distinguish three explanations. The hint may supply missing information; added context may redirect generation toward a solution the model could already produce; or implementation-level variation may change the output without any meaningful intervention effect. A pass/fail comparison alone cannot tell these cases apart.

HumanEval/117 makes the ambiguity concrete. The baseline program fails. The relevant hint, ``Consider how the current implementation treats uppercase and lowercase letters differently,'' produces a passing program. Yet a length-matched unrelated hint about the modulus operator also produces a pass, and at least one of eight unhinted samples passes as well. The relevant hint may still guide generation more effectively; the pass transition alone cannot tell us whether it supplied inaccessible information.

We study the distinction in code generation, where functional correctness can be evaluated by execution rather than by a language-model judge. We begin with short teacher-generated hints on HumanEval+ and MBPP+, then test successively smaller internal representations: a vector added to hidden activations, low-rank activation spaces, and short learned blocks of attention-memory vectors (virtual KV prefixes). At each stage, a control answers a narrower question. Replay measures pipeline instability. Unrelated hints measure generic prompt effects. Repeated no-hint sampling asks whether the rescued behavior already appears without a hint. Norm-matched random and shuffled interventions test task specificity. Held-out tasks test transfer beyond the data used to construct a representation. Size-matched prefixes test whether training beats an untrained object of the same size.

Controls narrow the interpretation at every stage. Unrelated hints rescue many failures, and most relevant-hint rescues also appear among eight unhinted samples. Relevant and unrelated hints induce nearly the same activation direction, whose full-benchmark deployment is associated with slightly more damage than rescue. Learned subspaces fit training deltas but have an imprecise held-out advantage over controls. Full context enables the synthetic procedures, whereas the tested KV-prefix objective fits its exemplars without consistent held-out transfer. These results are specific to the tested interventions; they do not establish a general impossibility.

Post-generation hidden states provide a separate positive result. A linear probe predicts functional correctness across benchmarks when all representation and hyperparameter choices are made on the source benchmark. Its top-one advantage over token confidence remains statistically unresolved, so we interpret the result as cross-benchmark correctness decodability rather than model self-knowledge.

Our contributions are:

\begin{enumerate}
\def\labelenumi{\arabic{enumi}.}
\tightlist
\item
  \textbf{A two-model behavioral audit of hint rescue.} We compare relevant hints with unrelated hints and no-hint best-of-eight sampling on 79 selected Qwen failures and independently repeat the pattern on 101 selected Phi failures; replay is measured in the primary Qwen study.
\item
  \textbf{A separation of geometric stability, behavioral change, and task specificity.} We show that a stable activation direction is associated with changed outputs without improving aggregate accuracy or carrying a detectable held-out task-specific residual.
\item
  \textbf{A controlled context-compression test.} We contrast full specifications with trained and size-matched virtual-KV prefixes on context-defined procedural tasks, while reporting the limits of the empirical no-context gate.
\item
  \textbf{A leakage-resistant correctness readout and reusable audit trail.} We select probe representations on source-benchmark grouped folds, test both transfer directions against text and surface baselines, and release task-level ledgers and the reanalysis script.
\end{enumerate}

Behavioral results cover Qwen and Phi; the mechanistic, compression, and correctness-readout analyses use Qwen only. The benchmarks contain short Python programs, and the prefix experiment tests one training objective. We therefore restrict our claims to these models, tasks, channels, and procedures.

\section{2. Operational Definitions}\label{operational-definitions}

We use the following operational definitions to keep behavioral access, intervention transfer, and correctness decoding distinct.

\textbf{Rescue.} A task is rescued by condition \(c\) when the baseline greedy completion fails the EvalPlus base and plus tests and the completion under \(c\) passes both.

\textbf{Sampled support at budget \(k\).} A task is in the model's sampled support at budget \(k\) when at least one of \(k\) no-hint samples passes. This is an empirical, budget-dependent definition; failure at finite \(k\) does not prove that success has zero probability.

\textbf{Semantic hint advantage.} The behavior attributable specifically to relevant content is the paired difference between relevant and attempt-, style-, length-, and seed-matched irrelevant hints. The present experiment includes unrelated hints but not every element of this matched estimand, so the observed 36-versus-19 difference is suggestive rather than a complete causal decomposition.

\textbf{Intervention transfer.} An activation object transfers when it is estimated without the evaluation task and improves held-out behavior relative to norm-, rank-, channel-, and compute-matched controls. High cosine stability or training-set energy capture alone is representational evidence, not transfer.

\textbf{Context-defined procedure.} A procedure is empirically context-dependent when repeated no-context and unrelated-context attempts rarely succeed but a full specification and worked examples succeed. We avoid the stronger phrase ``provably absent'' unless the task construction makes success information-theoretically impossible without a randomized secret.

\textbf{Correctness readout.} A probe is a readout when it predicts the execution label from hidden states. A successful readout establishes decodability under its train/test protocol. It does not by itself establish that the base model uses the feature during generation.

\section{3. Related Work}\label{related-work}

\subsection{3.1 Task vectors and activation interventions}\label{task-vectors-and-activation-interventions}

\citet{hendel2023taskvectors} and \citet{todd2024functionvectors} show that in-context demonstrations can induce compact task or function vectors whose addition reproduces behavior on controlled mappings. Activation Addition \citep{turner2023activation}, Contrastive Activation Addition \citep{panickssery2024caa}, and representation engineering \citep{zou2023representation} similarly use directions in activation space to steer high-level output properties. These results motivate our search for a compact representation of a helpful hint. Our setting differs in using long-form program generation, executable correctness, and task-level rescue and damage rather than only target-property change.

Causal efficacy is not sufficient for faithful interpretation. \citet{makelov2024subspace} demonstrate that a subspace patch can change behavior through a pathway that does not faithfully localize the hypothesized feature. We add complementary controls: split-half re-estimation, relevant-versus-irrelevant directions, a positive-control test of the intervention channel, held-out task transfer, matched random and shuffled interventions, and full-population net effects.

\subsection{3.2 Prompt sensitivity, placebos, and resampling}\label{prompt-sensitivity-placebos-and-resampling}

Self-consistency \citep{wang2023selfconsistency} established the practical value of sampling multiple reasoning trajectories. \citet{macar2025thought} argue that causal interpretation of reasoning also requires resampling rather than reliance on a single trajectory. \citet{mukherjee2024placebo} show that socio-demographic prompt effects can resemble responses to arbitrary placebo tokens, while \citet{kim2026noise} find that untrained random soft prompts can broaden early-token diversity and improve pass@N. \citet{luo2026missing} report that question-asking interventions depend strongly on self-consistency and can diagnose errors without reliably repairing them. These studies motivate our unrelated-hint, replay, and no-hint pass@8 controls. Code execution lets us measure whether each changed trajectory is functionally correct.

\subsection{3.3 Context and compact skill representations}\label{context-and-compact-skill-representations}

Gist tokens compress prompts using an attention-mask training scheme \citep{mu2023gist}, while task and function vectors compress demonstrated mappings into activations. \citet{petrov2024prefixlimits} characterize limitations of prompting and prefix tuning under specific architectural assumptions; \citet{petrov2024universal} also establish universal approximation for sufficiently large prefix constructions. The two results rule out a blanket theoretical claim that prefixes cannot represent new behavior.

Recent systems report positive skill storage using different substrates: Skill Neologisms learn soft vocabulary tokens \citep{berthon2026skill}, LatentSkill produces weight-space LoRA adapters \citep{yu2026latentskill}, and KV-Skill uses a learned interface to read external factorized operators \citep{han2026kvskill}. Our experiment is narrower. It tests whether a short, directly optimized virtual-KV prefix, trained with exemplar cross-entropy in a frozen model, preserves the behavior enabled by a textual specification and three examples.

\subsection{3.4 Correctness signals and code selection}\label{correctness-signals-and-code-selection}

Language-model self-evaluation and hidden-state probing have revealed correctness- or truthfulness-related signals in multiple domains \citep{kadavath2022mostlyknow, azaria2023internal, burns2023discovering}. \citet{orgad2024intrinsic} show that such signals need not transfer across datasets, making cross-benchmark evaluation important. More recent work probes arithmetic errors \citep{sun2025arithmetic}, chain-of-thought errors \citep{yuan2026hiddenerror}, and code correctness \citep{dicicco2026code, ribeiro2026robustness}. \citet{ashuach2026masked} further show that a model's own hidden states do not always outperform peer-model states, cautioning against claims of privileged self-knowledge.

Candidate selection for code has traditionally relied on generated tests, execution agreement, or program clustering \citep{chen2022codet, to2024functional}. UCoder incorporates internal probing into unsupervised code-model training \citep{wu2026ucoder}, while CASE uses task-grouped evaluation to show when hidden-state selection beats majority voting and exposes question-identity leakage in ordinary probe splits \citep{wang2026decodability}. We use source-benchmark training and target-benchmark evaluation, execution labels from EvalPlus, and within-task ranking. Unlike probe-only studies, we evaluate correctness readout as the endpoint of the same controlled pipeline that tests hint rescue, sampled accessibility, and activation transfer.

\section{4. Experimental Design}\label{experimental-design}

\subsection{4.1 Models, benchmarks, and execution}\label{models-benchmarks-and-execution}

The primary frozen student is Qwen2.5-3B-Instruct \citep{yang2024qwen25}. We independently repeat the behavioral hint and sampling stages with Phi-3.5-mini-instruct \citep{abdin2024phi, microsoft2024phi35}. Qwen2.5-Coder-7B-Instruct \citep{hui2024qwen25coder} serves only as an instrument for identifying teacher-solvable student failures and producing minimal hints; we do not treat it as an oracle. We evaluate the 164 HumanEval tasks \citep{chen2021humaneval} and 378 MBPP tasks \citep{austin2021program} augmented as HumanEval+ and MBPP+ by EvalPlus \citep{liu2023evalplus}. A program passes only when it satisfies both base and augmented tests. Unless stated otherwise, generation is greedy. Sampling experiments use temperature 0.8 and top-p 0.95. All activation, prefix, and correctness-readout results use the primary Qwen student.

The source runs used BF16, seed 42, batches of 12, left padding, and at most 512 new tokens on an NVIDIA GB10 Grace Blackwell system. The archived environment records Python 3.12.3, PyTorch 2.12.0+cu130, Transformers 5.9.0, EvalPlus 0.3.1, NumPy 2.2.6, SciPy 1.18.1, and scikit-learn 1.9.0. Each experimental stage has a timestamped run identifier, git commit, configuration, and summary in \texttt{results/runs.jsonl}. Because nominally identical greedy prompts sometimes change outcome under different batch compositions, we treat pipeline replay as an empirical control rather than assume determinism from the decoding rule.

The Qwen student passes 113 of 164 HumanEval+ tasks (68.9\%) and 249 of 378 MBPP+ tasks (65.9\%). The Phi student passes 108 (65.9\%) and 224 (59.3\%). The teacher passes 136 (82.9\%) and 268 (70.9\%), respectively. Intersecting teacher passes with student failures yields 79 selected Qwen failures (29 HumanEval+, 50 MBPP+) and 101 selected Phi failures (33, 68). These populations are selected specifically where the teacher passes and the student fails, so their rescue rates do not describe the full benchmarks; they also differ across students.

\subsection{4.2 Hint intervention and behavioral controls}\label{hint-intervention-and-behavioral-controls}

For each selected teacher-pass/student-fail task, the teacher produces an adaptive ladder of up to three minimal natural-language hints, typically no more than 23 words. The ladder proceeds from a light conceptual cue toward a more explicit cue and stops at the first pass. A task counts as rescued if any evaluated level passes. The unrelated control gives each selected task one level-1 hint generated for a different, already-passing task and chosen to be closest in token length to the relevant task's successful or last attempted hint. Other controls are no-hint best-of-eight sampling at temperature 0.8 and top-p 0.95, plus replay of the original greedy prompt under changed batch composition.

The relevant ladder offers one to three opportunities whereas the unrelated condition offers one; the sampling control changes both opportunity count and decoding rule. The implemented comparisons therefore answer whether these \emph{procedures} differ, not how much rescue is attributable specifically to semantic content.

\subsection{4.3 Activation capture and intervention}\label{activation-capture-and-intervention}

Hidden states are captured from the post-block residual stream at every layer. The anchor is the aligned token at the end of a suffix whose token IDs are identical across conditions. For task \(i\) and layer \(\ell\), the hint delta is

\[
\Delta_{i,\ell}=h^{\text{hint}}_{i,\ell}-h^{\text{base}}_{i,\ell}.
\]

The unit mean hint direction is

\[
g_\ell=\frac{\sum_i \Delta_{i,\ell}}
{\left\|\sum_i \Delta_{i,\ell}\right\|_2},
\]

and the reported per-delta energy fraction is the squared projection \((g_\ell^\top\Delta_{i,\ell})^2/\|\Delta_{i,\ell}\|_2^2\), summarized across tasks. Persistent injection applies \(h_{\ell,t}\leftarrow h_{\ell,t}+\alpha g_\ell\) at each decoding step \(t\) in the selected layer configuration. The \emph{full per-task oracle delta} is the unprojected \(\Delta_{i,\ell}\) from the hinted run known to rescue evaluation task \(i\); ``oracle'' denotes its dependence on that successful evaluation-task outcome.

We distinguish three tests. The first asks whether \(g_\ell\) is stable under split-half re-estimation. The second injects it persistently and evaluates both rescues and damage. The third patches the full oracle delta once at the aligned anchor as a positive-control test of single-position intervention. Random bases are matched to the learned intervention in rank and norm. Low-rank bases use three-fold held-out cross-fitting: within each split, the basis and layer/rank/strength configuration are fitted and selected using only the training tasks, then frozen for the held-out fold. Comparators are matched random bases and shuffled task deltas. The generic-direction deployment is exploratory: its mean direction, layer, and strength were not all cross-fitted before full-benchmark evaluation.

\subsection{4.4 Context-defined procedures and virtual-KV prefixes}\label{context-defined-procedures-and-virtual-kv-prefixes}

We construct four procedural families: balanced-ternary notation, an eight-operation stack language, ordered string rewriting, and a keyed codec. Each family contains three worked exemplars and six held-out problems with hidden executable tests. We evaluate no context, an unrelated procedure's context, and the full specification plus exemplars (approximately 800 tokens).

We then freeze the model and optimize virtual-KV prefixes of 2--16 tokens using exemplar cross-entropy. Each virtual token occupies 36,864 bytes in the tested implementation, so a prefix uses approximately 74--590 decimal kilobytes. We compare trained prefixes with untrained, random, shuffled, and size-matched controls. Low training loss verifies exemplar fit; only held-out execution measures transfer.

\subsection{4.5 Correctness readout and candidate selection}\label{correctness-readout-and-candidate-selection}

For each of 542 benchmark tasks, we draw eight samples, yielding 4,336 execution-labeled programs. HumanEval+ contributes 859 passing candidates among 1,312; MBPP+ contributes 1,875 among 3,024. We fit class-balanced, \(\ell_2\)-regularized logistic probes to unit-normalized hidden states. Candidate representations span layers \(\{8,14,20,26,32\}\), last-token or mean pooling, and \(C\in\{0.1,1,10\}\).

All representation, regularization, and score-combination choices are made without target-benchmark labels. Five-fold GroupKFold on source-benchmark task identifiers selects the layer, pooling rule, and \(C\), prioritizing source within-task AUROC, then source selection accuracy and pooled AUROC. The probe-plus-log-probability weight is chosen from \(\{0.25,0.5,1,2,4\}\) using the source out-of-fold scores. The selected model is then fit to the full source benchmark and evaluated once on the other benchmark. We report both HumanEval+\(\rightarrow\)MBPP+ and MBPP+\(\rightarrow\)HumanEval+.

We compare the readout with mean token log-probability, exact-match majority frequency, 23 engineered code-surface features, and a character 3--5-gram TF-IDF classifier. Pooled AUROC compares all labeled candidates; within-task AUROC compares correct and incorrect candidates from the same task, reducing the contribution of between-task difficulty. Candidate selection uses no test execution at inference time, but training uses execution labels and inference still requires eight generations and white-box hidden-state access. We exclude an exploratory analysis that selected representations on the evaluation benchmark; every reported probe result uses source-only model selection in \texttt{publication\_analysis.py}.

\subsection{4.6 Statistical reporting}\label{statistical-reporting}

Benchmark-task inference uses task as the unit, with generated samples and repeated interventions clustered within task. We use 1,000 task-bootstrap replicates for selector AUROC and accuracy intervals, 20,000 paired task-bootstrap replicates for subspace-control differences, exact McNemar tests for binary paired outcomes \citep{mcnemar1947sampling}, and 95\% Wilson intervals for individual binomial rates \citep{wilson1927probable}. Bootstrap intervals follow the task-resampling principle described by \citet{efron1993bootstrap}. Probe intervals condition on the realized source training set, source-selected model, and fixed eight-candidate target draw; they resample only target tasks. The random-subspace interval averages the five repeated interventions within each task and conditions on the realized folds and random bases. Procedure problems are nested within four families; the 24 problems do not support independent-family generalization. The subspace study is small (36 rescued tasks), so failure to reject a null is reported as \emph{no detected advantage}, not evidence of equivalence. The publication reanalysis was written after the exploratory runs and is therefore robustness analysis rather than preregistered confirmation.

\section{5. Most Hint Rescues Reappear Within Eight Unhinted Samples}\label{most-hint-rescues-reappear-within-eight-unhinted-samples}

Table 1 reports the principal behavioral controls on the selected teacher-pass/student-fail sets and the full Qwen benchmark.

\textbf{Table 1. Behavioral outcomes. Intervals are 95\% Wilson intervals. Replay rows apply only to the primary Qwen study.}

{\def\LTcaptype{none} 
\begin{longtable}[]{@{}
  >{\raggedright\arraybackslash}p{(\linewidth - 6\tabcolsep) * \real{0.2000}}
  >{\raggedleft\arraybackslash}p{(\linewidth - 6\tabcolsep) * \real{0.2667}}
  >{\raggedleft\arraybackslash}p{(\linewidth - 6\tabcolsep) * \real{0.2667}}
  >{\raggedleft\arraybackslash}p{(\linewidth - 6\tabcolsep) * \real{0.2667}}@{}}
\toprule\noalign{}
\begin{minipage}[b]{\linewidth}\raggedright
student and quantity
\end{minipage} & \begin{minipage}[b]{\linewidth}\raggedleft
count
\end{minipage} & \begin{minipage}[b]{\linewidth}\raggedleft
rate
\end{minipage} & \begin{minipage}[b]{\linewidth}\raggedleft
95\% CI
\end{minipage} \\
\midrule\noalign{}
\endhead
\bottomrule\noalign{}
\endlastfoot
Qwen: relevant-hint rescue & 36/79 & 45.6\% & {[}35.0, 56.5{]} \\
Qwen: unrelated-hint rescue & 19/79 & 24.1\% & {[}16.0, 34.5{]} \\
Qwen: no-hint best-of-8 success & 46/79 & 58.2\% & {[}47.2, 68.5{]} \\
Qwen: relevant rescues also solved by best-of-8 & 31/36 & 86.1\% & {[}71.3, 93.9{]} \\
Phi: relevant-hint rescue & 42/101 & 41.6\% & {[}32.5, 51.3{]} \\
Phi: unrelated-hint rescue & 17/101 & 16.8\% & {[}10.8, 25.3{]} \\
Phi: no-hint best-of-8 success & 57/101 & 56.4\% & {[}46.7, 65.7{]} \\
Phi: relevant rescues also solved by best-of-8 & 36/42 & 85.7\% & {[}72.2, 93.3{]} \\
Qwen: replay flips among all baseline failures & 5/180 & 2.8\% & {[}1.2, 6.3{]} \\
Qwen: replay flips among all baseline passes & 4/362 & 1.1\% & {[}0.4, 2.8{]} \\
\end{longtable}
}

\begin{figure}
\centering
\pandocbounded{\includegraphics[keepaspectratio,alt={Disjoint overlap patterns for the 79 selected Qwen failures. The relevant and unrelated conditions do not have matched attempt budgets; the figure emphasizes task identity rather than treating the three rates as independent.}]{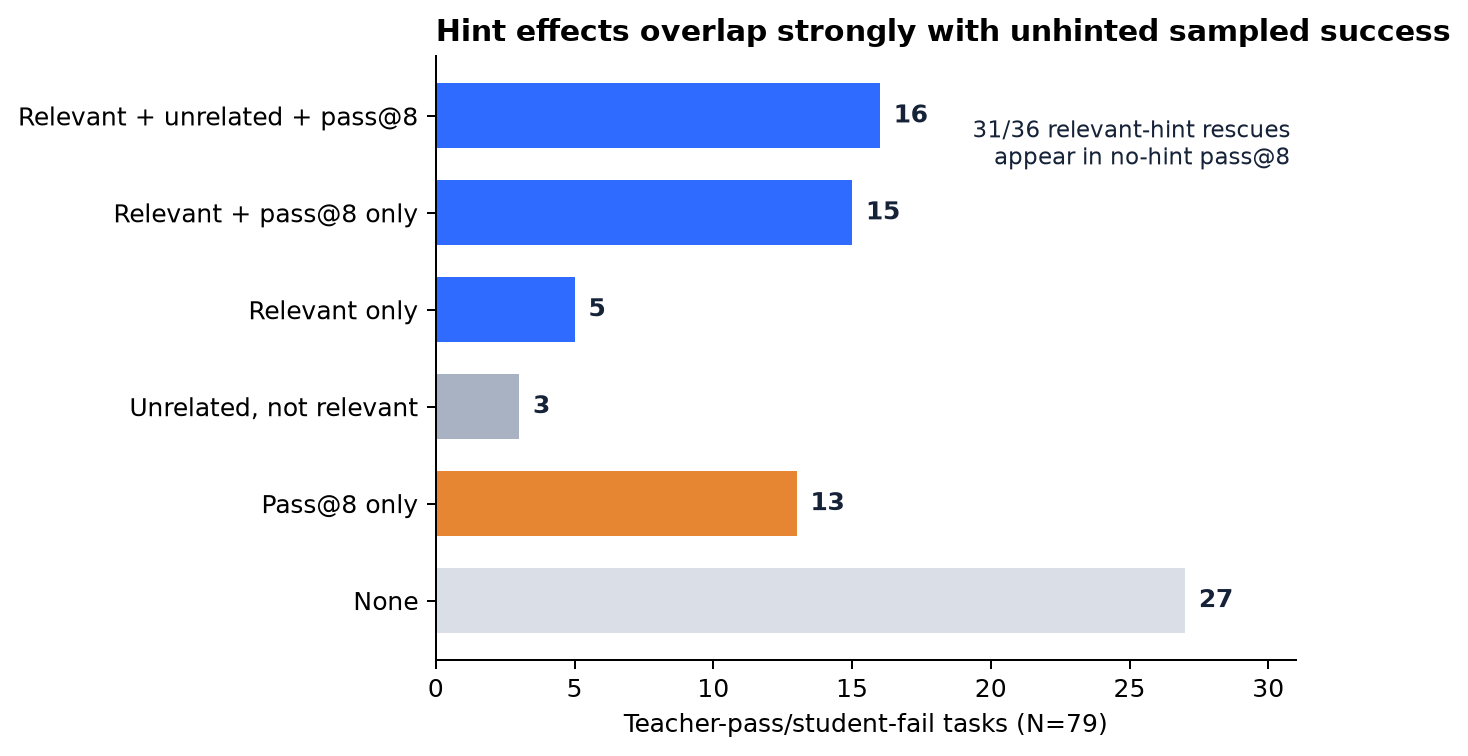}}
\caption{Disjoint overlap patterns for the 79 selected Qwen failures. The relevant and unrelated conditions do not have matched attempt budgets; the figure emphasizes task identity rather than treating the three rates as independent.}\label{fig-1}
\end{figure}

For Qwen, the observed ladder yields 36 rescues with relevant hints and 19 with unrelated hints. At the task level, 20 are rescued only by the relevant condition and 3 only by the unrelated condition (exact McNemar p = 0.00049). Phi shows 29 relevant-only and 4 unrelated-only tasks (p = 0.000011). These paired differences establish reproducible differences between the \emph{implemented procedures}, not semantic-content effects: attempt counts and stopping rules are not matched.

The sampling control changes the capability interpretation more sharply. Eight no-hint samples solve 46 of 79 selected Qwen tasks and 57 of 101 selected Phi tasks. They include 31 of 36 and 36 of 42 relevant-hint rescues, respectively. Thus, across both students, most observed rescues occur on tasks for which the unhinted model already produces a passing program within eight samples. This is evidence of accessibility under a modest sampling budget. It does not show that hinting and temperature sampling implement the same mechanism, nor does it explain the remaining five and six hint-only rescues.

On HumanEval+ as a whole, 77 of 164 tasks (47.0\%) are mixed across eight samples, while 61 are always solved and 26 are never solved. The large mixed region makes one-shot before/after comparisons especially fragile near the competence boundary.

Finally, nominal replay flips 5 baseline failures and 4 baseline passes without changing prompt content. A separate replay deployment used for the rescued-36 mechanistic analysis produces 6 of 36 passes (16.7\%). The latter is not a restriction of the full-benchmark replay run, in which 5 of 180 failures change to pass. The selected mechanistic population and separate deployment therefore have their own empirical replay reference.

\section{6. Mechanistic Results: Stability Is Not Specificity}\label{mechanistic-results-stability-is-not-specificity}

\subsection{6.1 Hinting produces a stable but largely generic direction}\label{hinting-produces-a-stable-but-largely-generic-direction}

Across layers, the mean hint-delta direction is extremely stable under the reported split-half re-estimation (cosine 0.992--0.996) and accounts for 35--63\% of the energy of individual hint deltas. This is a highly reproducible geometric pattern under the tested protocol. It is not, however, specific to useful hint content: mean directions estimated from relevant and unrelated hints have cosine similarity around 0.98 in early and middle layers.

An initially promising late-layer relevance direction appeared to separate relevant from irrelevant interventions on the 36 rescued tasks (27.8\% versus 16.7\%). The contrast appeared only after selecting a layer and the enriched 36-task subset; its failure to replicate points to selection noise rather than a stable semantic direction.

\begin{figure}[H]
\centering
\pandocbounded{\includegraphics[keepaspectratio,alt={The mean hint response is exceptionally stable under split-half re-estimation. Relevant and unrelated directions separate in late layers, but that late-layer contrast does not survive the causal robustness checks described below.}]{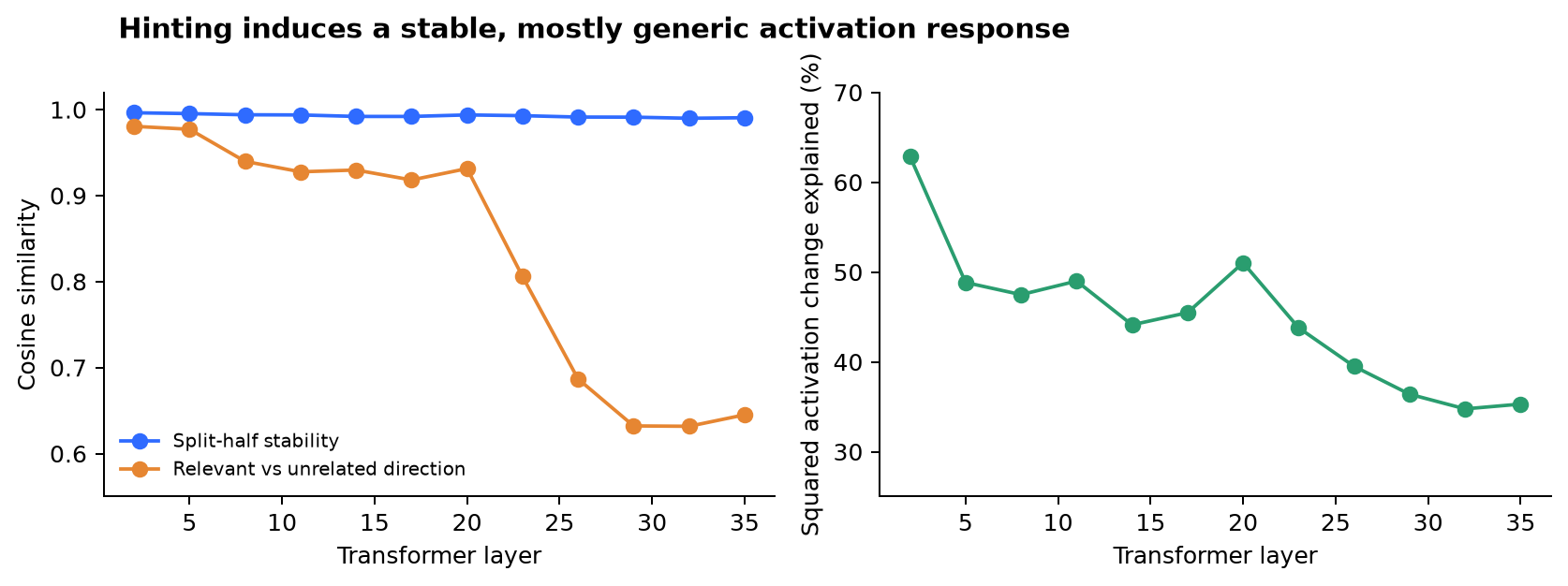}}
\caption{The mean hint response is exceptionally stable under split-half re-estimation. Relevant and unrelated directions separate in late layers, but that late-layer contrast does not survive the causal robustness checks described below.}\label{fig-2}
\end{figure}

\subsection{6.2 Persistent injection changes outcomes but not net accuracy}\label{persistent-injection-changes-outcomes-but-not-net-accuracy}

Persistent injection changes outcomes under the tested generation pipeline. On the 36 rescued tasks, 10 outputs pass under injection (27.8\%, 95\% Wilson CI {[}15.8, 44.0{]}) versus 6 under the separate replay deployment (16.7\%, {[}7.9, 31.9{]}). Because the execution paths are not documented as identical and deterministic, the difference is not identified causally. Full-benchmark deployment also measures damage to baseline passes.

\textbf{Table 2. Full-benchmark deployment of the generic direction. The net interval is a large-sample paired-difference interval based on the 14 beneficial and 18 harmful discordant pairs.}

{\def\LTcaptype{none} 
\begin{longtable}[]{@{}
  >{\raggedright\arraybackslash}p{(\linewidth - 6\tabcolsep) * \real{0.2000}}
  >{\raggedleft\arraybackslash}p{(\linewidth - 6\tabcolsep) * \real{0.2667}}
  >{\raggedleft\arraybackslash}p{(\linewidth - 6\tabcolsep) * \real{0.2667}}
  >{\raggedleft\arraybackslash}p{(\linewidth - 6\tabcolsep) * \real{0.2667}}@{}}
\toprule\noalign{}
\begin{minipage}[b]{\linewidth}\raggedright
transition
\end{minipage} & \begin{minipage}[b]{\linewidth}\raggedleft
count
\end{minipage} & \begin{minipage}[b]{\linewidth}\raggedleft
conditional rate
\end{minipage} & \begin{minipage}[b]{\linewidth}\raggedleft
95\% CI
\end{minipage} \\
\midrule\noalign{}
\endhead
\bottomrule\noalign{}
\endlastfoot
baseline fail to pass & 14/180 & 7.8\% & {[}4.7, 12.6{]} \\
baseline pass to fail & 18/362 & 5.0\% & {[}3.2, 7.7{]} \\
net passing programs & -4/542 & -0.74 percentage points & {[}-2.8, 1.3{]} percentage points* \\
\end{longtable}
}

\begin{figure}[H]
\centering
\includegraphics[width=0.78\linewidth,keepaspectratio,alt={Full-benchmark outcome transitions. Counts, rather than conditional percentages with different denominators, make the absence of net improvement visible.}]{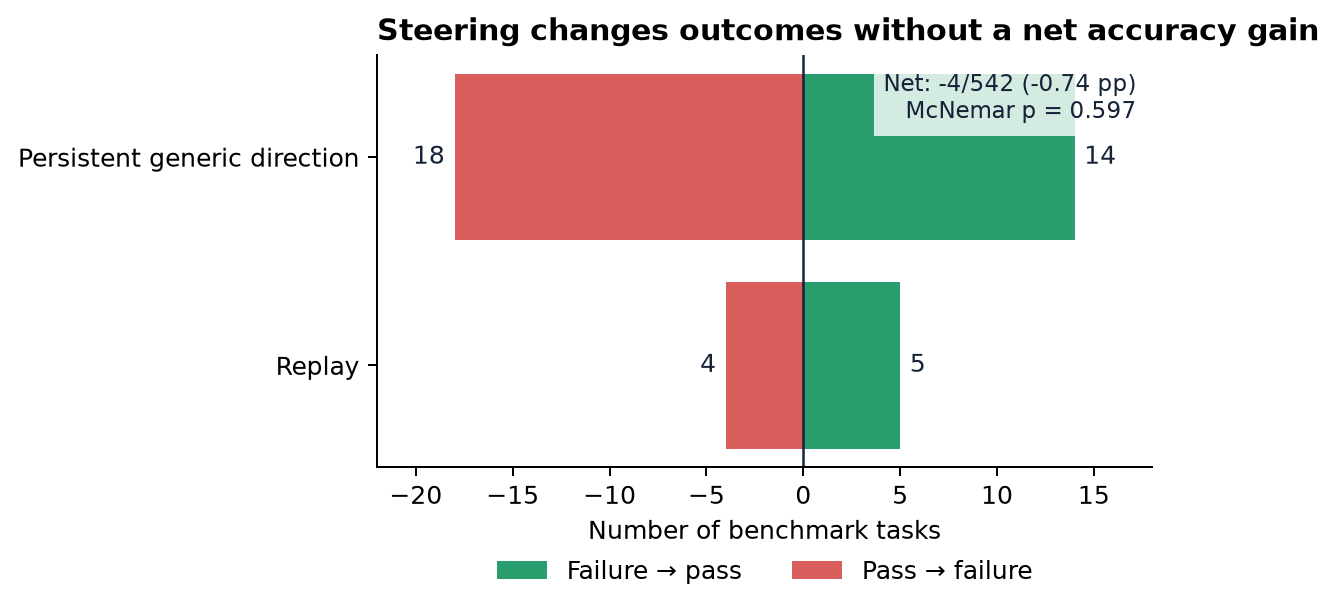}
\caption{Full-benchmark outcome transitions. Counts, rather than conditional percentages with different denominators, make the absence of net improvement visible.}\label{fig-3}
\end{figure}

The different conditional rates have different denominators. In absolute task counts, 18 passes are damaged and 14 failures are rescued. The transition imbalance is not statistically resolved (two-sided exact McNemar \(p=0.597\)); this does not establish equivalence. On the full benchmark, persistent injection is associated with both fail-to-pass and pass-to-fail transitions, with no detected net accuracy change. This establishes behavioral perturbation under the tested protocol, not a treatment effect isolated from pipeline variation.

\subsection{6.3 Single-position patching shows no detected effect in the tested channel}\label{single-position-patching-shows-no-detected-effect-in-the-tested-channel}

Patching the full per-task oracle rescue delta at the anchor position yields 6 of 36 passes (16.7\%, 95\% Wilson CI {[}7.9, 31.9{]}) for strengths \(\alpha\in\{1,2\}\), equal to the separate replay count. Persistent injection of the same delta changes more outcomes descriptively. No effect beyond replay is detected at this anchor and these strengths. Equality of the observed counts is not evidence of equivalence; instead, the experiment fails to validate this anchor as a positive-control channel.

\subsection{6.4 No held-out subspace advantage is detected}\label{no-held-out-subspace-advantage-is-detected}

Three-fold held-out cross-fitting over the 36 rescued tasks produces the following outcomes:

{\def\LTcaptype{none} 
\begin{longtable}[]{@{}
  >{\raggedright\arraybackslash}p{(\linewidth - 6\tabcolsep) * \real{0.2000}}
  >{\raggedleft\arraybackslash}p{(\linewidth - 6\tabcolsep) * \real{0.2667}}
  >{\raggedleft\arraybackslash}p{(\linewidth - 6\tabcolsep) * \real{0.2667}}
  >{\raggedleft\arraybackslash}p{(\linewidth - 6\tabcolsep) * \real{0.2667}}@{}}
\toprule\noalign{}
\begin{minipage}[b]{\linewidth}\raggedright
condition
\end{minipage} & \begin{minipage}[b]{\linewidth}\raggedleft
passes
\end{minipage} & \begin{minipage}[b]{\linewidth}\raggedleft
rate
\end{minipage} & \begin{minipage}[b]{\linewidth}\raggedleft
95\% Wilson CI
\end{minipage} \\
\midrule\noalign{}
\endhead
\bottomrule\noalign{}
\endlastfoot
learned low-rank subspace & 9/36 & 25.0\% & {[}13.8, 41.1{]} \\
replay & 6/36 & 16.7\% & {[}7.9, 31.9{]} \\
matched random subspaces & mean 5.8/36 across five repeats & 16.1\% & repeated within task \\
shuffled task deltas & 5/36 & 13.9\% & {[}6.1, 28.7{]} \\
\end{longtable}
}

Task-level paired comparisons yield +8.3 percentage points for learned minus replay (20,000-replicate task-bootstrap 95\% CI {[}-2.8, 19.4{]}; exact McNemar \(p=0.375\)), +8.9 points for learned minus the per-task mean of five matched-random runs ({[}-1.7, 20.6{]}), and +11.1 points for learned minus shuffled ({[}0.0, 25.0{]}; exact \(p=0.219\)). These intervals are wide and do not demonstrate equivalence, particularly at \(N=36\). The representational result points in the same direction: a training-estimated basis explains 64\% of training delta energy but approximately 9\% on held-out tasks. Residuals formed by removing the per-task placebo-direction component and leave-one-out ridge predictions show no detected improvement over their controls.

Under the tested ranks, layers, strengths, channel, and cross-validation protocol, the training-set subspace does not yield a detected task-specific transfer effect.

\section{7. Cross-Benchmark Correctness Readout}\label{cross-benchmark-correctness-readout}

The intervention experiments ask whether a compact state change can make the model execute a task-specific behavior. A complementary question is whether hidden states can rank behaviors the model has already generated.

Task-level candidate counts give a uniform-selection expectation of 107.4/164 (65.5\%) on HumanEval+; 102/164 is the realized first-draw count, not that expectation. MBPP+ contains 1,875 passing candidates among 3,024, for a uniform-selection expectation of 234.4/378 (62.0\%). Table 3 uses these denominators and source-only model selection.

\clearpage
\textbf{Table 3. Cross-benchmark candidate selection. Bootstrap intervals resample target tasks; paired tests compare the combined selector with mean log-probability.}

\begingroup
\footnotesize
\setlength{\tabcolsep}{2pt}
{\def\LTcaptype{none} 
\begin{longtable}[]{@{}
  >{\raggedright\arraybackslash}p{(\linewidth - 12\tabcolsep) * \real{0.1111}}
  >{\raggedleft\arraybackslash}p{(\linewidth - 12\tabcolsep) * \real{0.1481}}
  >{\raggedleft\arraybackslash}p{(\linewidth - 12\tabcolsep) * \real{0.1481}}
  >{\raggedleft\arraybackslash}p{(\linewidth - 12\tabcolsep) * \real{0.1481}}
  >{\raggedleft\arraybackslash}p{(\linewidth - 12\tabcolsep) * \real{0.1481}}
  >{\raggedleft\arraybackslash}p{(\linewidth - 12\tabcolsep) * \real{0.1481}}
  >{\raggedleft\arraybackslash}p{(\linewidth - 12\tabcolsep) * \real{0.1481}}@{}}
\toprule\noalign{}
\begin{minipage}[b]{\linewidth}\raggedright
target (probe source)
\end{minipage} & \begin{minipage}[b]{\linewidth}\raggedleft
uniform candidate, expected
\end{minipage} & \begin{minipage}[b]{\linewidth}\raggedleft
mean log-p
\end{minipage} & \begin{minipage}[b]{\linewidth}\raggedleft
hidden probe
\end{minipage} & \begin{minipage}[b]{\linewidth}\raggedleft
probe + log-p
\end{minipage} & \begin{minipage}[b]{\linewidth}\raggedleft
pass@8 oracle
\end{minipage} & \begin{minipage}[b]{\linewidth}\raggedleft
paired \(p\)
\end{minipage} \\
\midrule\noalign{}
\endhead
\bottomrule\noalign{}
\endlastfoot
MBPP+ \(\rightarrow\) HumanEval+ & 107.4/164 (65.5\%) & 113 (68.9\%) & 116 (70.7\%) & \textbf{122 (74.4\%)} {[}67.7, 81.1{]} & 138 (84.1\%) & 0.093 \\
HumanEval+ \(\rightarrow\) MBPP+ & 234.4/378 (62.0\%) & 240 (63.5\%) & 241 (63.8\%) & \textbf{244 (64.6\%)} {[}59.8, 69.0{]} & 281 (74.3\%) & 0.503 \\
\end{longtable}
}
\endgroup

\begin{figure}[H]
\centering
\pandocbounded{\includegraphics[keepaspectratio,alt={Target-task selection after all probe and combination choices are made on the source benchmark. Error bars are 95 percent task-bootstrap intervals. Dashed lines show the expected accuracy of a uniformly chosen candidate; dotted lines show the target pass@8 oracle.}]{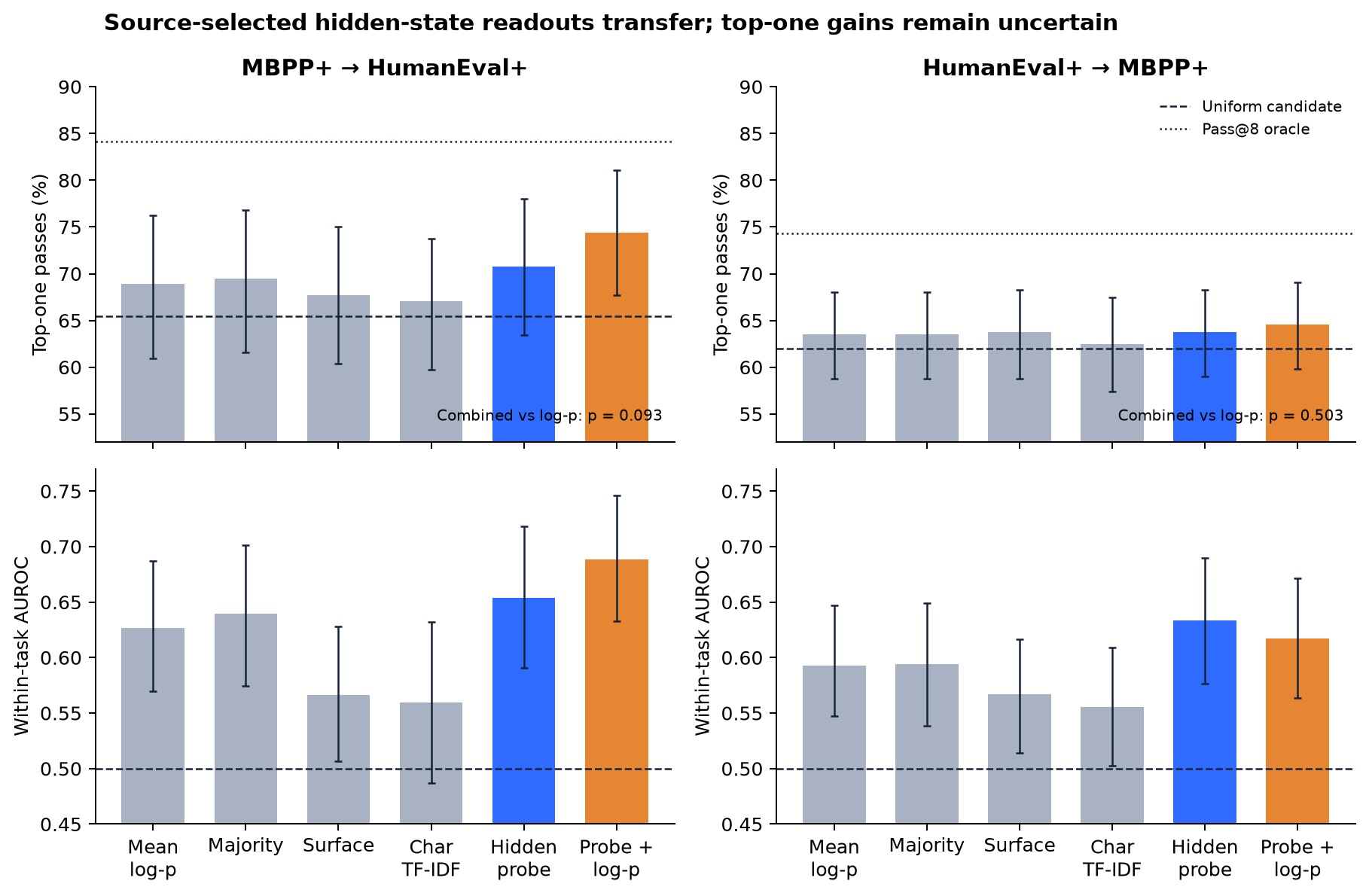}}
\caption{Target-task selection after all probe and combination choices are made on the source benchmark. Error bars are 95\% task-bootstrap intervals. Dashed lines show the expected accuracy of a uniformly chosen candidate; dotted lines show the target pass@8 oracle.}\label{fig-4}
\end{figure}

Both source benchmarks select layer 26, mean pooling, and \(C=10\). The hidden probe's pooled AUROC is 0.806 (task-bootstrap 95\% CI {[}0.750, 0.861{]}) on HumanEval+ and 0.780 ({[}0.742, 0.819{]}) on MBPP+. Within-task AUROC is lower but remains above chance: 0.654 ({[}0.591, 0.718{]}) and 0.634 ({[}0.577, 0.690{]}). For comparison, mean log-probability yields pooled/within-task AUROC 0.653/0.627 on HumanEval+ and 0.626/0.593 on MBPP+. The gap between pooled and within-task discrimination shows that task difficulty explains part, but not all, of the signal.

The surface controls narrow the interpretation further. A 23-feature syntax-and-length classifier reaches pooled/within-task AUROC 0.692/0.567 on HumanEval+ and 0.612/0.567 on MBPP+; a character TF-IDF classifier reaches 0.657/0.560 and 0.623/0.555. These surface baselines have lower point-estimate pooled AUROC than the hidden probe, although they carry substantial correctness information and do not exhaust possible textual confounds.

Top-one selection is less decisive than sample discrimination. On HumanEval+, the source-selected combination produces nine more passing selections than confidence alone: 16 tasks improve and 7 worsen, giving exact McNemar \(p=0.093\). On MBPP+, the difference is four tasks: 12 improve and 8 worsen, \(p=0.503\). Within the shared eight-candidate ledgers, first-draw counts are 102/164 and 237/378, whereas uniform-selection expectations are 107.375 and 234.375. Separate greedy runs solve 113/164 and 249/378; comparisons with those runs are descriptive cross-protocol counts, not the paired selector comparisons in Table 3. Exact-match majority selects 114 and 240 passing programs, respectively. These results support cross-benchmark correctness-related decodability; they do not yet establish a reliable top-one improvement over confidence or greedy decoding.

Probe training consumes execution labels for 4,336 programs. Inference avoids test execution but still requires eight generations and white-box hidden-state access. A post-generation probe may exploit code length, termination, syntax, memorized error signatures, or other properties not represented by our controls. The result is therefore a transferable \emph{readout}, not proof of privileged self-knowledge or causal use of the decoded feature during generation.

\section{8. Compressing Context-Defined Procedures}\label{compressing-context-defined-procedures}

The preceding benchmark tasks are often solved under sampling. To test a more explicit context gap, we construct four procedures defined by their supplied specifications and examples.

Across the four families, no-context generation solves at most 1 of 6 held-out problems per family despite 13 attempts per problem; an unrelated procedure's context also solves at most 1 of 6. The full specification plus three worked exemplars solves 22 of 24 held-out problems (91.7\%, 95\% Wilson CI {[}74.2, 97.7{]}). This establishes a large empirical context effect under execution. It does not prove that the pretrained model assigns zero probability to success without context.

Trained virtual-KV prefixes reach exemplar loss at or below 0.05 but solve only 5--11 of 24 held-out problems, overlapping the range of untrained, random, and shuffled controls. In a follow-up on the selected ordered-rewriting configuration with two virtual tokens, five perturbed-initialization seeds each solve the same 3 of 6 problems. Size-matched controls solve 2 of 6. The effective held-out unit remains six problems, not five seeds; the one-problem difference is insufficient evidence of procedure transfer.

This result is method-specific. The tested exemplar-cross-entropy virtual-KV scheme does not reproduce the full context on these four families, but other objectives, interfaces, lengths, and pretrained adapters may behave differently. Without a method-positive control or context-distillation objective, the experiment does not isolate a limit of the representation class from a limit of the training recipe. Full per-family results and controls appear in Appendix B.

\section{9. A Control Checklist for Capability-Transfer Claims}\label{a-control-checklist-for-capability-transfer-claims}

The experiments can be summarized by the interpretation each control changed.

{\def\LTcaptype{none} 
\begin{longtable}[]{@{}
  >{\raggedright\arraybackslash}p{(\linewidth - 6\tabcolsep) * \real{0.2500}}
  >{\raggedright\arraybackslash}p{(\linewidth - 6\tabcolsep) * \real{0.2500}}
  >{\raggedright\arraybackslash}p{(\linewidth - 6\tabcolsep) * \real{0.2500}}
  >{\raggedright\arraybackslash}p{(\linewidth - 6\tabcolsep) * \real{0.2500}}@{}}
\toprule\noalign{}
\begin{minipage}[b]{\linewidth}\raggedright
proposed inference
\end{minipage} & \begin{minipage}[b]{\linewidth}\raggedright
required control
\end{minipage} & \begin{minipage}[b]{\linewidth}\raggedright
observation here
\end{minipage} & \begin{minipage}[b]{\linewidth}\raggedright
supported interpretation
\end{minipage} \\
\midrule\noalign{}
\endhead
\bottomrule\noalign{}
\endlastfoot
a prompt caused a rescue & nominal replay on the same selected set & replay reaches 16.7\% on the rescued-36 subset & subtract or model the selected-set instability floor \\
the hint's content caused the rescue & attempt- and style-matched irrelevant hints & unrelated hints rescue 19/79 & generic conditioning is a substantial alternative; the semantic increment is not yet identified \\
rescue created a new capability & no-hint pass@k gate & pass@8 covers 31/36 hint rescues & most rescued tasks are already accessible within the tested sampling budget \\
a stable direction is semantically meaningful & split-half and relevant-vs-irrelevant estimation & stability is high, but relevant and irrelevant directions align & reproducibility does not imply content specificity \\
a steering vector improves the model & report rescue and damage on the full population & 14 rescues, 18 damages & the direction perturbs behavior without detected net benefit \\
a patched subspace is causally informative & positive-control channel validation & no effect beyond replay is detected for the full task-specific successful-hint difference at the tested anchor & the tested anchor was not validated as an effective causal channel \\
a learned subspace transfers & held-out cross-fitting plus task-clustered random and shuffled controls & +8.3 points vs replay, CI {[}-2.8, 19.4{]}; +8.9 vs random, CI {[}-1.7, 20.6{]} & estimates are too imprecise to establish transfer or equivalence \\
a trained virtual-KV prefix compresses a procedure & size-matched controls and held-out execution & 3/6 versus 2/6 in the selected-configuration follow-up & evidence is insufficient for transfer \\
a probe reveals correctness knowledge & source-only selection, cross-benchmark tests, and output-feature baselines & pooled AUROC 0.806/0.780; paired top-one \(p=0.093/0.503\) & correctness-related information is decodable; selection, cognitive, and causal claims remain open \\
\end{longtable}
}

Behavioral change, representational structure, intervention sensitivity, task specificity, generalization, and net utility are separate empirical claims. Each requires its own denominator and control.

\section{10. Discussion}\label{discussion}

\subsection{10.1 What the experiments support}\label{what-the-experiments-support}

First, a before-and-after hint rescue is a poor proxy for newly acquired capability on these benchmarks. Most relevant-hint rescues occur on tasks the unhinted student solves within eight samples, and unrelated hints also change many outcomes. Relevant content may contribute, but the present ladder does not isolate that contribution from unequal intervention opportunities or other prompt differences.

Second, a representation can be real and reproducible without supporting the intended mechanistic interpretation. The generic hint direction is exceptionally stable and is behaviorally associated with changed outputs under persistent injection. Its alignment across relevant and irrelevant prompts and its negative full-benchmark net change make ``task-specific capability vector'' an unsupported label.

Third, the negative intervention results are method-specific. Single-position patching shows no detected effect at the tested anchor, while persistent injection changes outcomes. The learned subspace shows no detected held-out advantage, but the sample is small. The KV-prefix objective transfers little on the synthetic procedures, but other prefix lengths, objectives, pretrained interfaces, and weight-space methods may behave differently.

Fourth, the post-generation readout transfers in both benchmark directions and has higher point-estimate AUROC than the tested confidence and surface baselines. Its top-one gain over confidence is not statistically resolved, and its observed MBPP+ count is below that of the separate greedy run. This readout does not turn a failure into a success; it can only help choose a success already present among the candidates.

\subsection{10.2 Relation to concurrent work}\label{relation-to-concurrent-work}

Our results sit between positive work on task/function vectors and recent cautions about interpreting activation interventions. Function-vector studies demonstrate compact causal representations on controlled input-output mappings, while subspace-patching work shows that behavioral effects need not imply faithful localization. Our code experiments extend the control question to long-form generation with executable labels, but do not invalidate positive results in other task families.

Recent prompting and resampling work makes the behavioral controls especially important. Socio-demographic placebo experiments show that arbitrary prompt tokens can perturb outputs even when their content is irrelevant. Random soft prompts can broaden early token distributions and improve pass@N without learned content. Work on thought-branch resampling likewise argues that a single generated trajectory is insufficient for causal interpretation. Our setting differs in using functional code execution and in tracing the same ambiguity from prompts into activation interventions, but these results reinforce the need to compare content-bearing interventions with compute-matched perturbations and resampling.

The prefix result must also be read against both sides of the expressivity literature. Petrov et al.~characterize restrictions of prompting and prefix tuning under particular architectural conditions, while their later universal-approximation result shows that sufficiently large constructions can be expressive in principle. Recent Skill Neologism, LatentSkill, and KV-Skill results report positive task storage using different substrates or learned interfaces. Our finding is therefore an empirical failure of one compact virtual-KV scheme on four context-defined procedures, not a theorem about prefix-class methods.

Finally, hidden-state correctness detection is now a crowded area. Prior and concurrent work reports truthfulness, reasoning-error, arithmetic-error, and code-correctness signals, including robustness and confound analyses. CASE shows that question-grouped evaluation is necessary to avoid question-identity leakage when comparing hidden-state selection with voting. UCoder uses internal probing as part of unsupervised code-model training, and peer-probe comparisons show that a model's own states need not carry privileged correctness information in every domain. Our distinct evidence is the combination of EvalPlus execution labels, cross-benchmark training and testing, within-task comparisons, and direct candidate-selection utility in the same controlled study. We do not claim the first discovery of correctness decodability.

\section{11. Limitations}\label{limitations}

\textbf{Model scope.} The behavioral hint/resampling pattern repeats in two student architectures, but the intervention, prefix, and probe results cover only Qwen2.5-3B-Instruct. This does not establish scale or broad architecture generality.

\textbf{Benchmark scope and contamination.} HumanEval+ and MBPP+ contain short Python problems and are widely used. Augmented tests improve outcome validity but do not remove possible benchmark exposure. A contamination-resistant, time-split, or newly authored benchmark would strengthen external validity.

\textbf{Adaptive hint ladder.} Up to three relevant-hint levels provide multiple chances for rescue, while the length-matched unrelated condition provides one. The no-hint pass@8 arm also uses a different decoding rule. Hints were generated during the experimental pipeline rather than frozen in a preregistered artifact before student evaluation.

\textbf{Pipeline nondeterminism.} Outcome changes under nominal replay complicate every small causal effect, especially after selection on rescued tasks. The archived runs do not use single-example batches, deterministic kernels, or identical execution paths across every intervention.

\textbf{Exploratory multiplicity and power.} Layer, rank, strength, pooling, and direction searches create multiple comparisons. Held-out cross-fitting protects the reported subspace test tasks, but the causal subset contains only 36 tasks. No prespecified smallest effect of interest or equivalence interval is reported.

\textbf{Intervention scope.} The point patch uses one anchor and two strengths. The persistent intervention and the low-rank construction test particular residual-stream channels. Null results cannot be generalized to other positions, heads, components, or learned interfaces.

\textbf{Synthetic-procedure gate.} Thirteen failed attempts do not prove absence, and some constructed rules may overlap with patterns in pretraining. The study does not use randomized secrets or per-instance rule permutations, and the virtual-KV experiment has one objective without a demonstrated method-positive control.

\textbf{Probe confounds and cost.} Post-generation hidden states may encode surface artifacts correlated with correctness. The probe requires white-box access, eight generations, and execution-labeled training data. We add length/syntax and character TF-IDF baselines, task-clustered intervals, and paired selector tests, but omit stronger code encoders, static analyzers, peer-model states, residualized covariates, and external calibration tests.

\section{12. Conclusion}\label{conclusion}

Across two small models, relevant hints help under the implemented conditions, but no-hint sampling covers most successful rescues. In the primary Qwen mechanistic study, apparent capability transfer becomes substantially narrower under direct controls. A stable hint direction is associated with changed outputs, yet it is shared across relevant and irrelevant prompts and provides no detected net accuracy gain. Learned activation subspaces have positive but imprecise held-out estimates; compact virtual-KV prefixes remain far below the full-context condition. Post-generation hidden states have higher point-estimate AUROC than the tested baselines in both transfer directions, while their incremental top-one selection value remains preliminary.

A capability-transfer claim should report replay, matched placebos, the no-hint sampling boundary, causal-channel validation, held-out comparisons, and damage alongside rescue. Reporting these controls distinguishes interventions that merely change outputs from those that improve accuracy or transfer task-specific behavior.

\section{Data and Artifact Availability}\label{data-and-artifact-availability}

A compact reproducibility artifact has been prepared for public archival deposit. It contains the full experimental source, configuration, dependency lock, run registry, task-level ledgers for baselines, hints, sampling gates, interventions, and procedures, 4,336 raw sampled programs and execution labels, cached probe representations, the independent Phi behavioral replication, the publication reanalysis, and all figures. Running

\begin{Shaded}
\begin{Highlighting}[]
\ExtensionTok{python}\NormalTok{ publication\_analysis.py }\AttributeTok{{-}{-}root}\NormalTok{ . }\AttributeTok{{-}{-}output}\NormalTok{ publication\_analysis}
\end{Highlighting}
\end{Shaded}

recreates the corrected statistics and figures without model inference. The compact review bundle omits approximately 5.9 GB of raw per-layer activation tensors and trained checkpoints retained in the authors' full archive; these are required to rerun intervention construction, but not to reproduce any reported table, interval, paired test, or figure from the released ledgers and cached features. \texttt{ARTIFACT\_MANIFEST.md} maps every claim to its source files and records the limitation.

\section{Appendix A. Evidence and Reproduction Map}\label{appendix-a.-evidence-and-reproduction-map}

\begingroup
\footnotesize
{\def\LTcaptype{none} 
\begin{longtable}[]{@{}
  >{\raggedright\arraybackslash}p{(\linewidth - 4\tabcolsep) * \real{0.3333}}
  >{\raggedright\arraybackslash}p{(\linewidth - 4\tabcolsep) * \real{0.3333}}
  >{\raggedright\arraybackslash}p{(\linewidth - 4\tabcolsep) * \real{0.3333}}@{}}
\toprule\noalign{}
\begin{minipage}[b]{\linewidth}\raggedright
claim family
\end{minipage} & \begin{minipage}[b]{\linewidth}\raggedright
primary archived evidence
\end{minipage} & \begin{minipage}[b]{\linewidth}\raggedright
publication output
\end{minipage} \\
\midrule\noalign{}
\endhead
\bottomrule\noalign{}
\endlastfoot
Qwen baselines and selected failures & \path{results/baseline_summary_*.json}, \path{results/baseline_results_*.jsonl} & Table 1 denominators \\
relevant and unrelated hints & \path{results/hints_*.jsonl}, \path{results/controls_*.json}, \path{results/rescued_*.json} & Table 1, Figure 1 \\
no-hint best-of-eight & \path{results/gate_results_humaneval+mbpp.json} & Table 1, Figure 1 \\
Phi behavioral replication & \path{replication_phi/results/*}, \path{replication_phi/behavioral_summary.json} & Table 1 \\
geometry and causal interventions & \path{results/xray*.json}, \path{results/causal_cv_results_humaneval+mbpp.json}, \path{results/deploy_results_humaneval+mbpp.json} & Table 2, Figures 2--3 \\
candidate selection & \path{data/samples_*.jsonl}, \path{cache/selfeat_*.npz}, \path{publication_analysis/publication_statistics.json} & Table 3, Figure 4 \\
context-defined procedures & \path{results/skillgaps.json}, \path{results/capsule_distill.json}, \path{results/capsule_replication.json} & Appendix B \\
execution and software provenance & \path{results/runs.jsonl}, \path{config.yaml}, \path{requirements.txt}, \path{artifacts/system_info.txt}, \path{logs/} & Methods and artifact audit \\
\end{longtable}
}
\endgroup

The main Qwen registry contains timestamped stages for smoke validation, baselines, hints, activation capture, subspace estimation, held-out causal evaluation, intervention deployment, resampling, candidate generation, probe analysis, procedure gating, prefix distillation, and replication. Each record contains the stage's git commit. The Phi directory has a separate registry and source configuration. The corrected analysis uses seed 20260828 for bootstrap resampling and does not execute generated code or call a language model.

\section{Appendix B. Virtual-KV Results by Family}\label{appendix-b.-virtual-kv-results-by-family}

The four context-defined families are balanced-ternary notation (TRN), an eight-operation stack language (GSL), ordered string rewriting (CRW), and a keyed codec (KZE). Table B1 reports whether any of eight generated programs passes each held-out problem; all cells are counts out of six.

\textbf{Table B1. Any-of-eight held-out success for trained virtual-KV prefixes and controls.}

{\def\LTcaptype{none} 
\begin{longtable}[]{@{}
  >{\raggedright\arraybackslash}p{(\linewidth - 16\tabcolsep) * \real{0.0857}}
  >{\raggedleft\arraybackslash}p{(\linewidth - 16\tabcolsep) * \real{0.1143}}
  >{\raggedleft\arraybackslash}p{(\linewidth - 16\tabcolsep) * \real{0.1143}}
  >{\raggedleft\arraybackslash}p{(\linewidth - 16\tabcolsep) * \real{0.1143}}
  >{\raggedleft\arraybackslash}p{(\linewidth - 16\tabcolsep) * \real{0.1143}}
  >{\raggedleft\arraybackslash}p{(\linewidth - 16\tabcolsep) * \real{0.1143}}
  >{\raggedleft\arraybackslash}p{(\linewidth - 16\tabcolsep) * \real{0.1143}}
  >{\raggedleft\arraybackslash}p{(\linewidth - 16\tabcolsep) * \real{0.1143}}
  >{\raggedleft\arraybackslash}p{(\linewidth - 16\tabcolsep) * \real{0.1143}}@{}}
\toprule\noalign{}
\begin{minipage}[b]{\linewidth}\raggedright
family
\end{minipage} & \begin{minipage}[b]{\linewidth}\raggedleft
k=2
\end{minipage} & \begin{minipage}[b]{\linewidth}\raggedleft
k=4
\end{minipage} & \begin{minipage}[b]{\linewidth}\raggedleft
k=8
\end{minipage} & \begin{minipage}[b]{\linewidth}\raggedleft
k=16
\end{minipage} & \begin{minipage}[b]{\linewidth}\raggedleft
untrained context-init k=8
\end{minipage} & \begin{minipage}[b]{\linewidth}\raggedleft
untrained random k=8
\end{minipage} & \begin{minipage}[b]{\linewidth}\raggedleft
trained random-init k=8
\end{minipage} & \begin{minipage}[b]{\linewidth}\raggedleft
shuffled prefix
\end{minipage} \\
\midrule\noalign{}
\endhead
\bottomrule\noalign{}
\endlastfoot
TRN & 1 & 0 & 1 & 1 & 1 & 0 & 0 & 1 \\
GSL & 3 & 1 & 1 & 3 & 1 & 1 & 1 & 1 \\
CRW & 4 & 1 & 2 & 2 & 2 & 1 & 1 & 1 \\
KZE & 0 & 0 & 0 & 0 & 0 & 0 & 0 & 0 \\
\end{longtable}
}

\renewcommand{\thefigure}{B\arabic{figure}}
\setcounter{figure}{0}
\begin{figure}
\centering
\pandocbounded{\includegraphics[keepaspectratio,alt={Held-out any-of-eight success by trained virtual-KV prefix length. Each virtual token occupies 36 KiB in the tested implementation. The dashed line is the per-family mean for the full textual context (22/24 overall); controls are reported in Table B1.}]{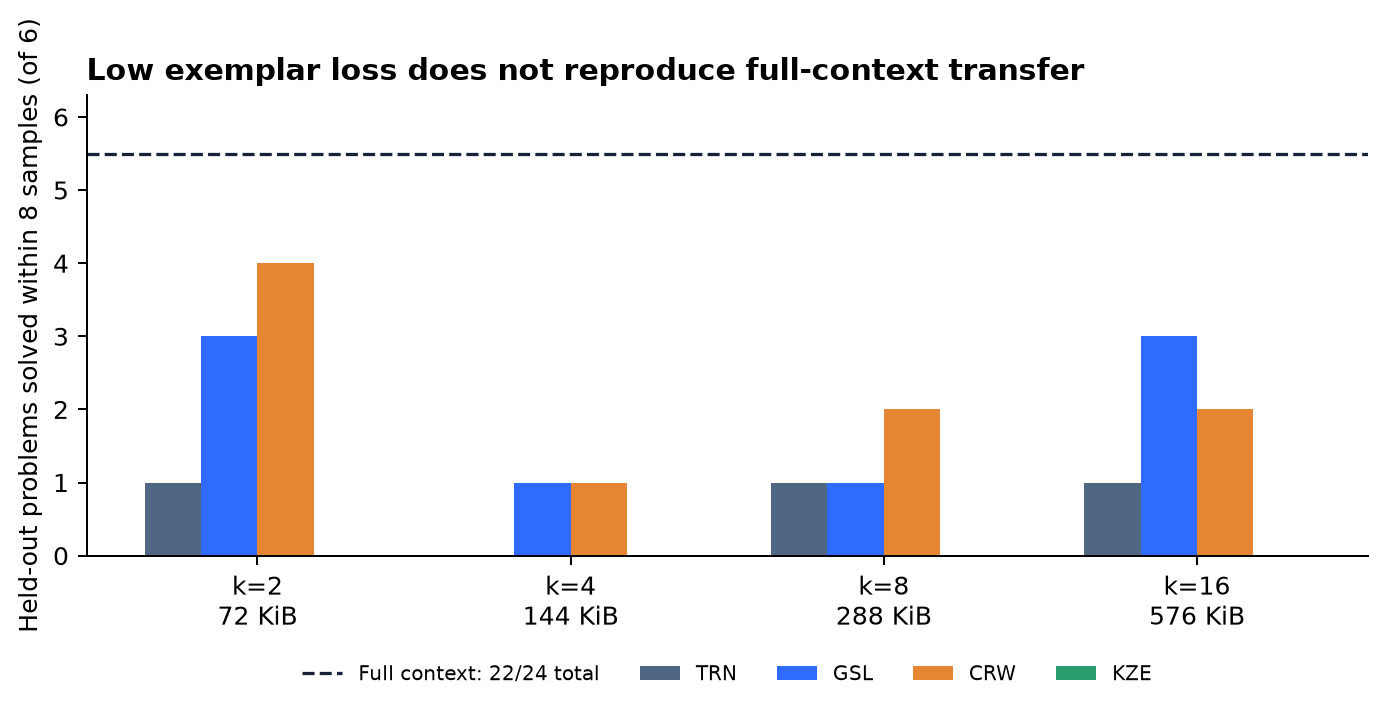}}
\caption{Held-out any-of-eight success by trained virtual-KV prefix length. Each virtual token occupies 36 KiB in the tested implementation. The dashed line is the per-family mean for the full textual context (22/24 overall); controls are reported in Table B1.}\label{fig-b1}
\end{figure}

The selected CRW k=2 follow-up uses fresh test cases and five perturbed initializations. Every run solves the same three of six cases; size-matched controls solve two. This is reproducible problem-level overlap, not five independent six-problem replications.

\section{Appendix C. Design Boundaries at a Glance}\label{appendix-c.-design-boundaries-at-a-glance}

{\def\LTcaptype{none} 
\begin{longtable}[]{@{}
  >{\raggedright\arraybackslash}p{(\linewidth - 6\tabcolsep) * \real{0.2500}}
  >{\raggedright\arraybackslash}p{(\linewidth - 6\tabcolsep) * \real{0.2500}}
  >{\raggedright\arraybackslash}p{(\linewidth - 6\tabcolsep) * \real{0.2500}}
  >{\raggedright\arraybackslash}p{(\linewidth - 6\tabcolsep) * \real{0.2500}}@{}}
\toprule\noalign{}
\begin{minipage}[b]{\linewidth}\raggedright
comparison
\end{minipage} & \begin{minipage}[b]{\linewidth}\raggedright
shared unit
\end{minipage} & \begin{minipage}[b]{\linewidth}\raggedright
opportunity/decoding match
\end{minipage} & \begin{minipage}[b]{\linewidth}\raggedright
identified interpretation
\end{minipage} \\
\midrule\noalign{}
\endhead
\bottomrule\noalign{}
\endlastfoot
relevant vs unrelated hint & same selected task & no: adaptive one-to-three vs one greedy attempt & difference between implemented procedures; semantic increment not isolated \\
relevant hint vs no-hint pass@8 & same selected task & no: greedy hint ladder vs eight stochastic draws & overlap with observed unhinted accessibility, not mechanistic equivalence \\
generic steering vs replay & benchmark task & same reported decoding settings, but batch/kernel paths not deterministic & associated outcome transitions beyond an empirical replay reference \\
learned subspace vs replay/shuffled/random & rescued task held out by fold & rank/norm matched; random interval conditions on realized bases & imprecise held-out advantage estimate, not equivalence \\
hidden readout vs confidence & same target task and eight candidates & yes for candidate pool; training/model-selection uncertainty conditioned on & paired top-one comparison on the realized candidate ledger \\
\end{longtable}
}

\bibliography{references.bib}

\end{document}